\documentstyle[a4,epsf]{article}
\begin{document}
\parindent = 36pt
\noindent{\small OG 4.1.8}

\baselineskip=18pt
\centerline{\Large \bf THE DETECTION OF TEV GAMMA RAYS}
\centerline{\Large \bf FROM CRAB USING }
\centerline{\Large \bf THE TELESCOPE ARRAY PROTOTYPE}

\baselineskip=15pt
\medskip

\centerline{\bf (The Telescope Array Collaboration)}

\medskip
\baselineskip =14pt

\noindent S.Aiso$^3$, 
M.Chikawa$^9$, 
Y.Hayashi$^1$, 
N.Hayashida, 
K.Hibino$^4$, 
H.Hirasawa, 
K.Honda$^5$, 

\noindent
N.Hotta$^{14}$, 
N.Inoue$^3$,
F.Ishikawa, 
N.Ito$^1$, 
S.Kabe$^{10}$, 
F.Kajino$^2$, 
T.Kashiwagi$^4$, 

\noindent
S.Kawakami$^1$, 
Y.Kawasaki$^1$, 
N.Kawasumi$^5$, 
H.Kitamura$^8$, 
K.Kuramochi$^{17}$, 
K.Kurata$^2$, 

\noindent
E.Kusano$^3$, 
H.Lafoux, 
E.C.Loh$^6$, 
\noindent
Y.Matsubara$^{12}$, 
T.Matsuyama$^1$, 
Y.Mizumoto$^{11}$, 

\noindent
K.Mizutani$^3$, 
M.Nagano, 
D.Nishikawa, 
J.Nishimura$^4$,
M.Nishizawa$^{16}$, 
T.Ouchi, 
H.Ohoka, 

\noindent
M.Ohnishi, 
To.Saito$^{15}$, 
N.Sakaki, 
M.Sakata$^2$, 
K.Sakumoto$^{18}$
M.Sasaki, 
M.Sasano$^1$, 

\noindent
H.Shimodaira,
A.Shiomi$^3$,
P.Sokolsky$^6$, 
T.Sugiyama$^2$, 
T.Takahashi$^1$, 
S.Taylor$^6$, 
M.Teshima, 

\noindent
I.Tsushima$^5$,
Y.Uchihori$^7$, 
T.Yamamoto, 
Y.Yamamoto$^2$, 
S.Yoshida, 
H.Yoshii$^{13}$, 
and T.Yuda

\medskip

{\small
\baselineskip=12pt
\noindent{\it Institute for Cosmic Ray Research, University of Tokyo, Tokyo 188, Japan.} \par
\noindent{\it $^1$Department of Physics, Osaka City University, Osaka 558, Japan.} \par
\noindent{\it $^2$Department of Physics, Konan University, Kobe 658, Japan.} \par
\noindent{\it $^3$Department of Physics, Saitama University, Urawa 338, Japan.} \par
\noindent{\it $^4$Department of Engineering, Kanawaga University, Yokohama 221, Japan.} \par
\noindent{\it $^5$Faculty of Education, Yamanashi University, Kofu 400, Japan}\par
\noindent{\it $^6$Department of Physics, University of Utah, USA}\par
\noindent{\it $^7$The 3rd Research Group, National Institute of Radiological Sciences, Chiba 263, Japan}\par
\noindent{\it $^8$Department of Physics, Kobe University, Kobe 657, Japan}\par
\noindent{\it $^9$Research Institute for Science and Technology, Kinki University, Osaka 577, Japan}\par
\noindent{\it $^{10}$National Laboratory of High Energy Physics, Tsukuba 305 ,Japan}\par
\noindent{\it $^{11}$National Astronomical Observatory, Tokyo 181, Japan}\par
\noindent{\it $^{12}$Solar-Terrestrial Environment Laboratory, Nagoya University, Nagoya 464-1, Japan}\par
\noindent{\it $^{13}$Faculty of General Education, Ehime University, Ehime 790, Japan}\par
\noindent{\it $^{14}$Faculty of Education, Utsunomiya University, Utsunomiya 321, Japan}\par
\noindent{\it $^{15}$Tokyo Metropolitan Colleage of Aeronautical Engineering, Tokyo 116, Japan}\par
\noindent{\it $^{16}$National Center for Science Information System, Tokyo 112, Japan}\par
\noindent{\it $^{17}$Informational Communication, KOKUSAI Junior College, Tokyo 165, Japan}\par
\noindent{\it $^{18}$Faculty of Science and Technology, Meisei University, Tokyo 191, Japan}\par

}

\baselineskip=15pt
\medskip

\centerline{\bf ABSTRACT}

The Telescope Array prototype detectors were installed at Akeno Observatory
and at the Utah Fly's Eye site. Using these detectors, we have observed the
Crab Nebula and AGN's since the end of 1995. The successful detections
of TeV gamma rays from Crab Nebula and Mkn501 are reported.

\medskip

\noindent{\bf 1.INTRODUCTION}

The physics objectives of the Telescope Array are the studies of
Ultra High Energy Cosmic Rays and Very High Energy Gamma Rays
(Teshima {\it et al.} 1992).
The development work is going on at Akeno and at the Fly's Eye site in Utah,
where the prototype detectors have been installed.
The seven telescopes array is under construction at Utah, and
three are in operation. 
With these prototype detectors, we have carried out the observations
of the Crab nebula and AGN's in the TeV gamma ray mode.
We will report the Crab observation results from Akeno and Utah, and
Mkn501 flares observed at Utah site during March, April and May in 1997.

\medskip

\baselineskip = 15pt

\noindent{\bf 2.EXPERIMENT}

The seven telescope array, telescope array prototype, 
is under construction at the Fly's Eye site at Cedar Mountain in Utah.
The altitude is 1,600 m above sea level, and
the geographical position is $40.33^{\circ}$ N and $113.02^{\circ}$ W.
Seven telescopes will be arranged 
at the grid of a regular hexagon with a separation of 70m.
Three telescopes are already in operation. 
The array will be in full operation in this autumn.

Telescopes are protected in the daytime by shelters in order to
avoid dust, rain and light. 
Telescopes and shelters can be controlled from the counting room 
located near the center of the array.
Each telescope has a 3 m diameter dish which
consists of nineteen hexagonal segment mirrors.
The effective mirror area in each telescope is 6 $m^2$ and 
the reflectivity is about 90 \% at 400 nm.
The 256 channel cameras with 0.25 degree pixels are mounted on the focal 
plane
of the telescopes.  We employed multi anode photomultipliers (MAPMT)
having 4 pixels(2$\times$2). 
In each MAPMT, a 4 mm thick BG3 filter is attached on the PMT
window. The filter is used as a light guide as well as an optical filter.
The filter has a transmission of 90 \% for the wavelength 
from 300 to 450 nm.

\begin{figure}[h]
\begin{center}
{{\epsfxsize6cm\epsfbox{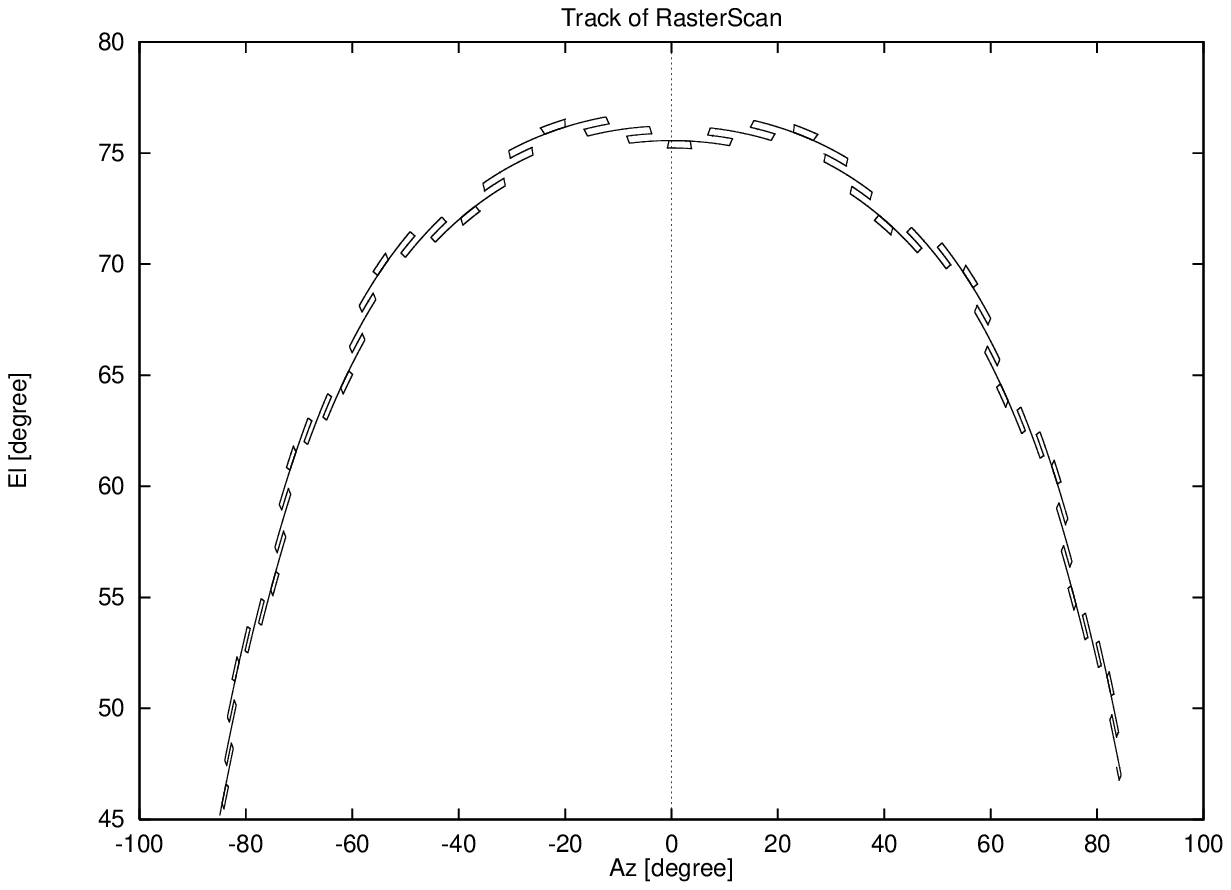}} \ \vspace{0.5cm} {\epsfxsize7cm\epsfbox{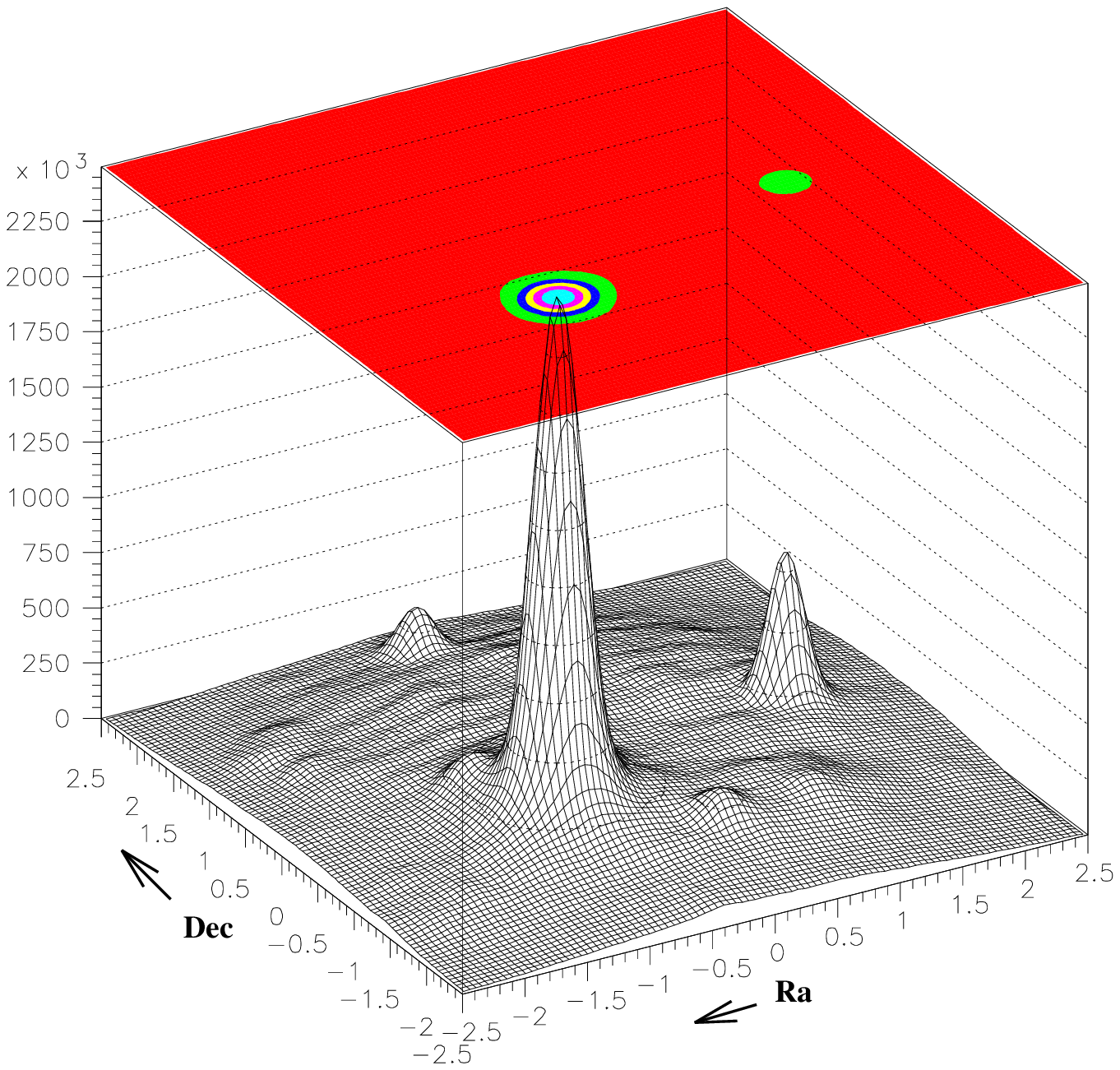}}}
\caption{
	{\bf Left:}The trajectory of the telescope tracking center in 
        the raster scan mode. 
	{\bf Right:} Examples of the star map around Crab Nebula. 
	These maps can be made
        by accumulating background photons on the equatorial 
        coordinate. From the bright star peak position, 
	we can calibrate the absolute direction of the telescopes
        with an accuracy of $0.03^\circ$.
	}
\label{fig:star}
\end{center}
\end{figure}

The signals from the PMT are amplified just behind the PMT camera and 
then are
fed to the data acquisition system mounted on the telescope driving system.
The pulse height and the pulse timing of the signals from PMT are
measured in each event.
The rise time of signals is about 10nsec and
the TDC values are digitized with 1 nsec accuracy. 
With the TDC information, the background photon noise 
can be efficiently eliminated in the image processing.
The threshold level of discriminators corresponds to four
photo-electrons and the single PMT tube rate is 5KHz.
We require four hit channels for triggering the events.
The resultant cosmic ray event rate in each telescope becomes 
about 4 $\sim$ 5 Hz.
With the simulation analysis, the energy threshold for the detectable
gamma rays is found to be 600 GeV.
The new data acquisition system based on C-MOS FADC is under development,
and will be installed this winter. With this new system, the inter telescope 
triggers can be made and a lower threshold energy will be achieved.

Alt-azimuth mounts are employed for the driving system of the telescope.
The mount is controlled by the network local computer VxWorks.
The rotation angles are read by the encoder with an accuracy of
0.001 degrees every 0.1 second.
The local computer calculates the target position and the velocity, 
and then these values are compared with the actual telescope's position 
and velocity. The difference between them are minimized by a feed back loop
in the computer software. The typical tracking accuracy 
is $\sim 0.005^\circ$, however, for small zenith angle regions, 
the velocity of the azimuth axis increases rapidly and 
the tracking error in the azimuth axis becomes larger. 
We therefore defined the dead space of the zenith angle less than 
$4^\circ$ for the tracking.

The new star tracking method called raster scan is employed in our experiment.
During a raster scan the center of the field of view scans
the square region of $1^\circ/cos(dec) \times 1^\circ$ in right ascension and
declination coordinate centered on the target.
One raster scan cycle is 48 minutes.
The advantages of this tracking method are as follows:
1) obtain the larger FOV than the actual FOV,
2) reduce the systematics due to the non-uniformities of PMT's gains and
due to the star light, 
3) more efficient observations because the target is always inside the 
FOV, and
4) calibration of the absolute direction of the telescope is possible using
bright stars.

\bigskip

\noindent{\bf 3.ANALYSIS}

The pedestal of ADC is measured by an artificial trigger 
every 10 seconds during the observations.
The pedestal values as a function of time for each channel are used
in the analysis.
The image processing from the raw data is required to reject the night
sky background photons. The images which show the concentration in the geometry
and in the time are selected using the timing information
(the software time window is $\pm 20nsec$).
After these image processing, the usual imaging analysis (Weekes et al. 1989) 
is carried out.
The selection of the events with the image parameters is 
used to enrich the gamma showers with the following conditions;
$0.05^\circ \le {\bf WIDTH} \le 0.15^\circ$, 
$0.15^\circ \le {\bf LENGTH} \le 0.45^\circ$ and 
$0.4^\circ \le {\bf DISTANCE} \le 1.3^\circ$. 
These selection parameters were determined to maximize 
the $ Q= S/\sqrt{N}$ by a monte carlo simulation that includes every 
experimental condition: the telescope optics, the geometry of mirrors, 
the shade of the PMT clusters, the PMT alignment, 
and the effect of the raster scan.
With this selection, gamma rays are selected with an efficiency of 50\%.



\begin{figure}[h]
\centerline{{\epsfxsize7.8cm\epsfbox{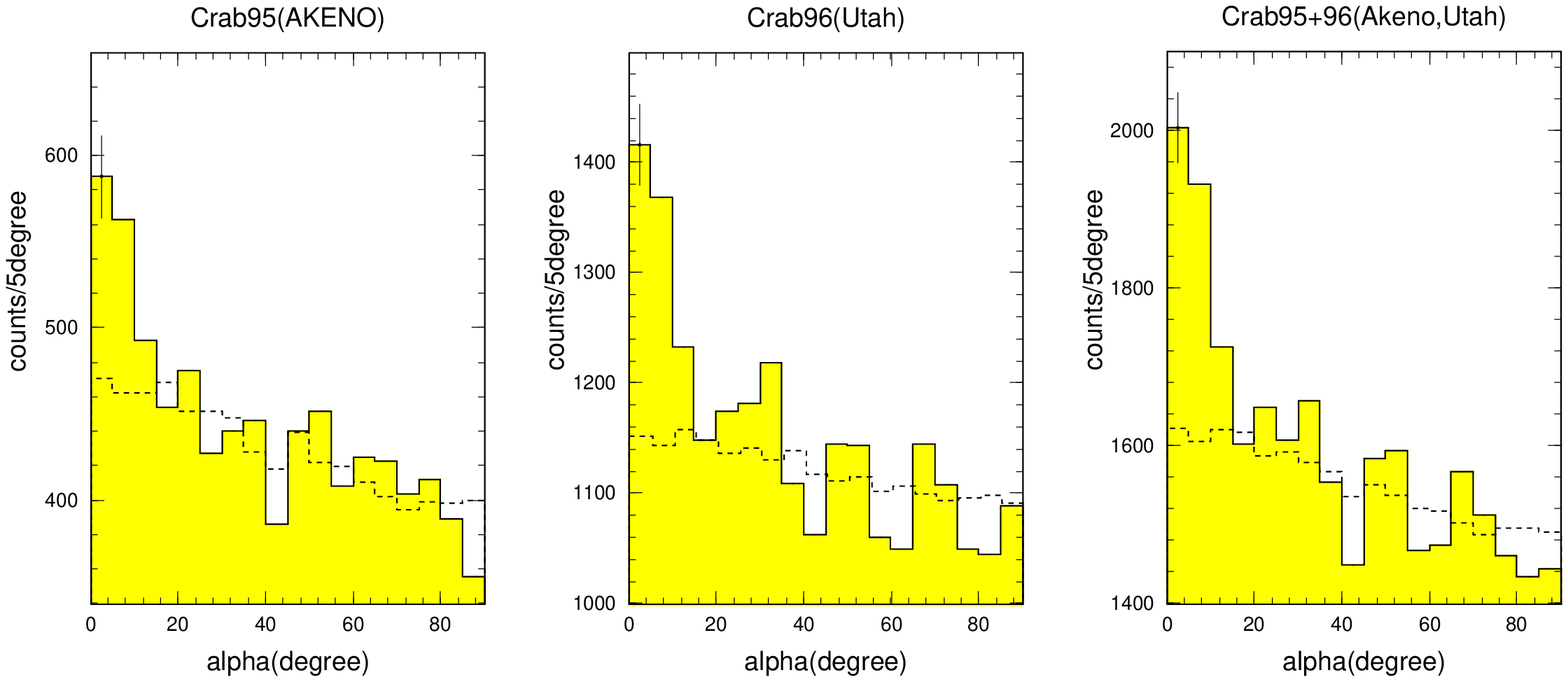}}}
\vspace{-0.3cm}
\caption{The alpha distribution after the image cut for Crab data.  
{\bf Left}: Crab observation carried out at Akeno observatory with two
telescopes in 1995, {\bf center}: Crab ovservation at Utah with two
telescopes in 1996,{\bf  right}: Combined data of Akeno95 and Utah96. }
\label{fig:crab}
\end{figure}

\noindent{\bf 3.1 CRAB NEBULA}

We have analysed Crab Nebula data taken at Akeno from Oct. 1995 to
Dec. 1995 and at Utah from Nov. 1996 to Feb. 1997.
We have used only clear sky night data for analysis. 
The observation times amount to 40 hours and 60 hours 
at Akeno and at Utah respectively.
The numbers of total events are about 311,000 and 1,070,000 
after the firt stage analysis which selects cerenkov events and determines
the orientation and image parameters.
Figure \ref{fig:crab} shows the alpha plot after the image parameter cut.
The total significance exceeds 10 $\sigma$.
From the analysis using Monte Carlo simulation,
the effective area after the image selection is calculated
as 12,000 $m^2$.
The flux of gamma rays from Crab Nebula is estimated to be 
0.93$\times10^{-11}$[/$cm^2$/sec] at 1 TeV using the data taken at Utah site.


\noindent{\bf 3.2 MKN501}

\begin{figure}[h]
\centerline{\hspace{0.5cm} {\epsfxsize7cm\epsfbox{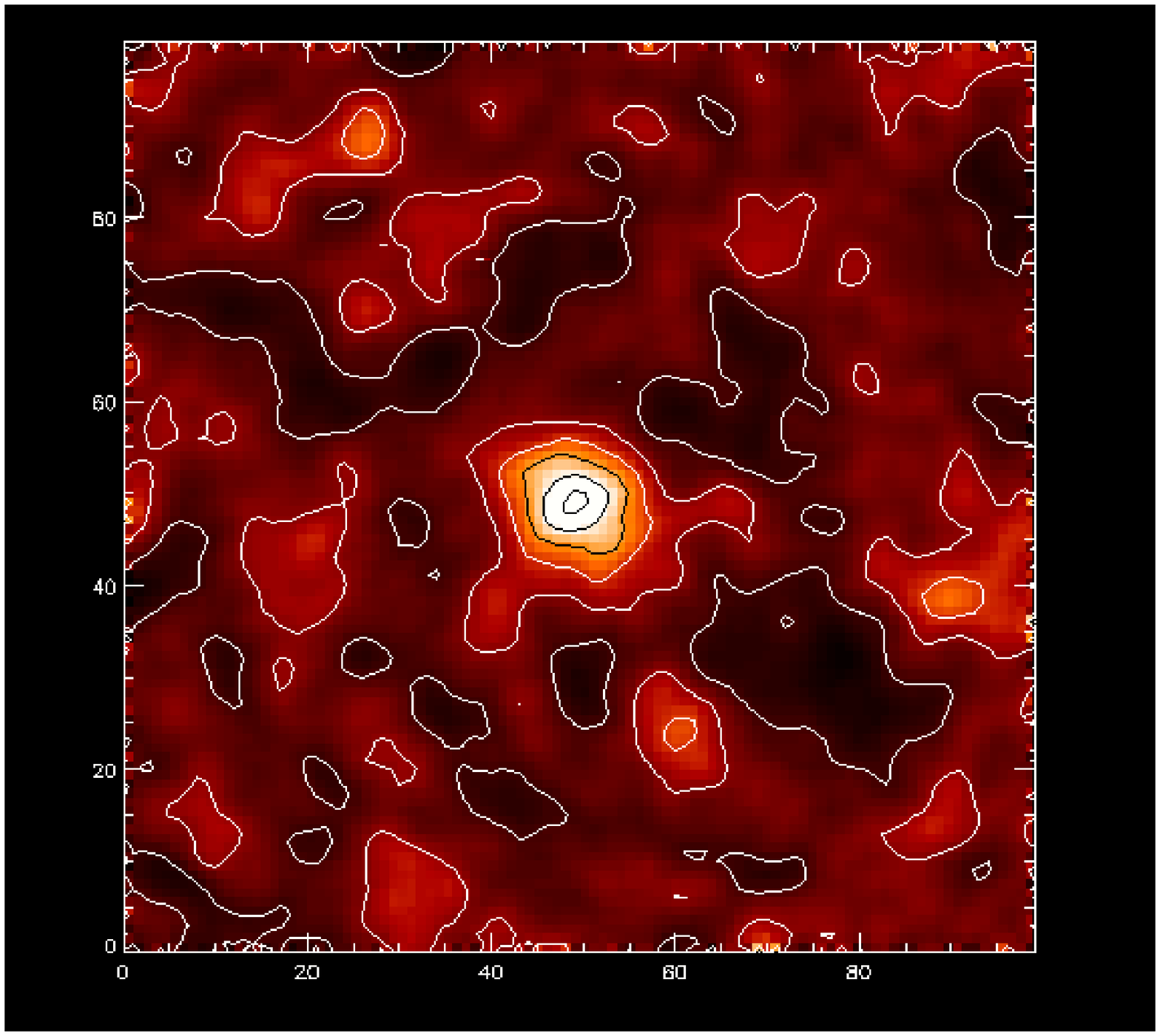}}, \ \ {\epsfxsize7cm\epsfbox{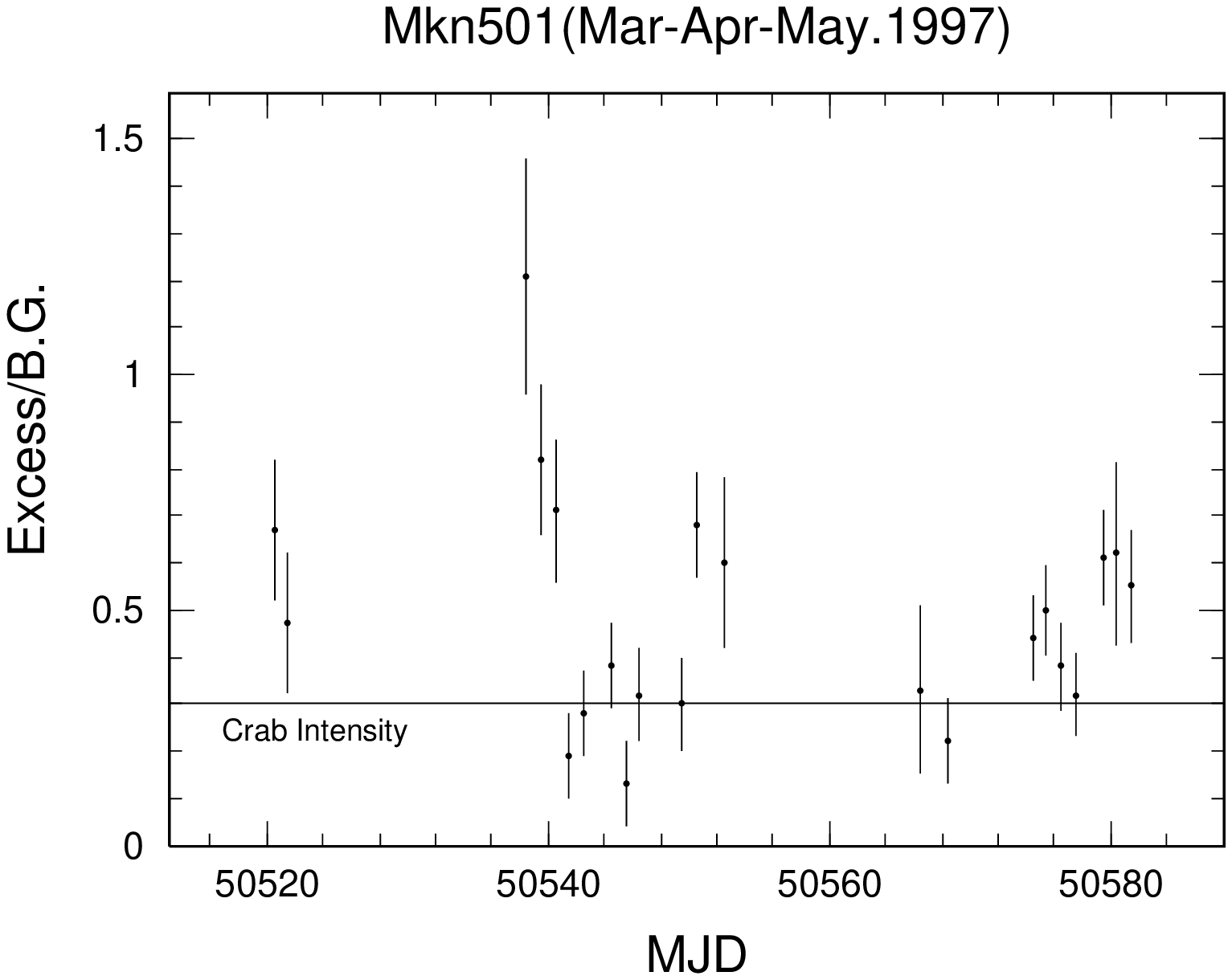}}}
\caption{The strong gamma ray flares from Mkn501 were detected by
Telescope Array prototype detector in March, April and May 1997.
{\bf Left}: The map of excess events in $\alpha \le 10^\circ$. 
The Mkn501 is located at the center of the map. 
The FOV of the map is $5^\circ \times 5^\circ$.
{\bf Right}: The daily excess rates. The horizontal line
shows the intensity of gamma rays from the Crab Nebula.}
\label{fig:mkn501}
\end{figure}

During March, April and May 1997, we have observed 
strong flares from Mkn501.
The gamma ray intensities during these period
exceed the Crab flux and sometimes became stronger 
than the Crab by a factor of 2 - 4.
The significance map and daily intensities are shown
in figure \ref{fig:mkn501}. The intensities have changed with 
the time scale of a few days. We have
tried to find very short time variation within one day data,
however, we could not find any such burst.

\medskip

\noindent{\bf 4.CONCLUSION}

The seven telescopes array is under construction at Utah Fly's Eye site. 
The construction will be completed in this summer.
At present, three telescopes are in operation. 
The gamma rays from Crab Nebula and from Mkn501 were successfully detected
with high statistics.
Gamma ray flares from Mkn501 have been detected, and the time profiles
are under the investigation. We have seen the variation of intensities
within a scale of a few days but failed to see any short bursts.
The flares of Mkn501 are still continuing 
while we are writing this manuscript.  Further details of our results will be
reported in the conference or elsewhere.

\medskip

\noindent{\bf ACKNOWLEDGEMENTS}

This work is supported in part by the Grants-in-Aid 
for Scientific Research (Grants \#07247102 and \#08041096) 
from the Ministry of Education, Science and Culture.
The authors would like to thank Mrs.B.Jones, Mr.R.Smith, Mr.S.Thomas 
and Mr.A.Larsen for their technical supports, and the people at Dugway
for the help of observations.

\noindent{\bf REFERENCE}\par
\parindent =0pt
T.C.Weekes {\it et al.} 1989, Ap.J. {\bf 342} p379.\par
M.Teshima {\it et al.} 1992, Proceedings of Towards a Major Atmospheric 
Cerenkov Detector for TeV astro/particle Physics, edited by P.Fleury and G.Vacanti, p255.\par
\end{document}